%% file: bitripartiteToMPECS2020_arXiv3.tex
\documentclass[acmsmall,nonacm]{acmart}
\usepackage{graphicx,subfig}
\usepackage{amsthm}
\usepackage{amsmath}
\usepackage{mathtools}
\usepackage{amsthm}
\usepackage{tikz}
\usepackage{bm}
\graphicspath{{./figs/}}
\newcommand{\eg}{\emph{e.g.}}
\newcommand{\ie}{\emph{i.e.}}

\newtheorem{claim}{Claim}

\AtBeginDocument{%
  \providecommand\BibTeX{{%
    \normalfont B\kern-0.5em{\scshape i\kern-0.25em b}\kern-0.8em\TeX}}}

\setcopyright{none}
\renewcommand\footnotetextcopyrightpermission[1]{}
\begin{document}

\title{On the Capacity Region of Bipartite and Tripartite Entanglement Switching}

\author{Gayane Vardoyan}
\affiliation{%
  \institution{University of Massachusetts, Amherst}
}
\email{gvardoyan@cs.umass.edu}
\author{Philippe Nain}
\affiliation{%
  \institution{Inria, France}
 }
  \email{philippe.nain@inria.fr}

\author{Saikat Guha}
\affiliation{%
  \institution{The University of Arizona}
}
  \email{saikat@optics.arizona.edu}

\author{Don Towsley}
\affiliation{%
   \institution{University of Massachusetts, Amherst}
}
 \email{towsley@cs.umass.edu}


\begin{abstract}
We study a quantum entanglement distribution switch serving a set of users in a star topology with equal-length links.  
The quantum switch, much like a quantum repeater, can perform entanglement swapping to extend entanglement across longer distances. Additionally, the switch is equipped with entanglement switching logic, enabling it to implement switching policies to better serve the needs of the network.
In this work, the function of the switch is to create bipartite or tripartite entangled states among users at the highest possible rates at a fixed ratio. Using Markov chains, we model a set of randomized switching policies. Discovering that some are better than others, we present analytical results for the case where the switch stores one qubit per user, and find that the best policies outperform a time division multiplexing (TDM) policy for sharing the switch between bipartite and tripartite state generation.  This performance improvement decreases as the number of users grows.  The model is easily augmented to study the capacity region in the presence of quantum state decoherence and associated cut-off times for qubit storage, obtaining similar results.  Moreover, decoherence-associated quantum storage cut-off times appear to have little effect on capacity in our identical-link system. We also study a smaller class of policies when the switch stores two qubits per user.
\end{abstract}

\ccsdesc[500]{Networks~Network performance evaluation}
\ccsdesc[500]{Hardware~Quantum communication and cryptography}

\keywords{quantum switch; entanglement distribution; Markov chain}


\maketitle
\section{Introduction}
\input{Introduction_extend}
\section{Background and Related Work}
\label{sec:backrel}
\input{BackgroundRelatedWork}
\section{System Description and Assumptions}
\label{sec:probform}
\input{ProbForm_extend}
\section{System with Per-Link Buffer Size One}
\label{sec:B1}
\input{B1Descr_extend}
\subsection{Numerical Results}
\input{NumericalB1_extend}
\subsection{Analysis}
\label{sec:B1analysis}
\input{Analysis_extend}
\section{System with Per-Link Buffer Size Two}
\label{sec:B2}
\input{B2Descr_extend}
\section{Modeling Decoherence and Quantum Storage Cut-Off Time}
\label{sec:Decoherence}
\input{Decoherence_extend}

\section{Conclusion}
\label{sec:conclusion}
\input{Conclusion_extend}

\begin{acks}
The work was supported in part by the National Science Foundation under grants CNS-1617437 and CNS-1955834. It was also supported by the grant ERC 1941583. Saikat Guha acknowledges support of an NSF subaward of a Yale University led project, grant number 1640959.
\end{acks}

\bibliographystyle{abbrv}
\bibliography{qnet}

\end{document}

%% file: Introduction_extend.tex
Multi-qubit entangled states are fundamental building blocks for quantum computation, sensing, and security. Consequently there is a need for a quantum network that can generate such entanglement on demand between pairs and groups of users~\cite{pirandola2019end,pant2019routing,Schoute2016-ir,Van_Meter2014-vz,dahlberg2019link}. In this paper, we study the performance of the simplest multi-user network, a star-topology quantum switch connecting $k$ users, where each user is connected to the switch via a separate link. Bipartite, two-qubit maximally-entangled states, \emph{i.e.}, Bell pairs (or EPR states) are generated at a constant rate across each link, with the qubits stored at local quantum memories at each end of the links. As these link-level entangled states start appearing, the switch uses two-qubit Bell-state measurements (BSM) between pairs of locally-held qubits and three-qubit Greenberger-Horne-Zeilinger (GHZ) basis measurements between triples of locally-held qubits to provide two-qubit and three-qubit entanglement to pairs and triples of users, respectively~\cite{nielsenchuang}. The capacity of such a switch to provide these two types of entanglement to the users depends on the switching mechanism, the number of quantum memories and their decoherence rates (assuming that quantum state decoherence imposes a qubit storage cut-off policy implemented at the switch), and the total number of links or users. 

We define the bipartite and tripartite capacities of the switch as the highest possible rate at which the device can generate Bell pairs and 3-qubit GHZ states via entanglement swapping, respectively, under a specified entanglement switching policy. For a given number of end nodes and quantum memories, the set of all switching policies define the capacity region of the switch.
In this paper, we study the capacity region when the switch can store either $B=1$ or $B=2$ qubits for each link at any given time. 
The number of quantum memories available to a link is referred to as its buffer size. We consider a simple time division multiplexing (TDM) policy between the two types of entangled states, along with a class of randomized switching policies.
In a TDM policy, the switch performs BSMs a fixed fraction of the time, and 3-way GHZ basis measurements for the remaining portion of the time.
When properly configured, the randomized switching policies provide higher capacities than TDM.  However, the relative difference between the two sets of policies goes to zero as $k \rightarrow \infty$. We also observe that increasing the number of memories from one to two increases capacity but that the increase diminishes as $k$ increases. 
Since the locally stored qubits at each end of the link are subject to a noise process that reduces the entanglement between the two qubits, we also explore the effect that decoherence has on capacity.
Specifically, we assume that the switch implements a cut-off policy for qubit storage to cope with finite coherence time of quantum memories. Throughout this manuscript, we often implicitly refer to storage cut-off times when referring to studying a system with quantum state decoherence. In the cases of $B=1$ with and without decoherence, we have simple closed form expressions for capacity whereas for the case of $B=2$, we provide a partial analysis but our results are mainly numerical.

The remainder of this paper is organized as follows: in Section \ref{sec:backrel}, we provide relevant background and related work. In Section \ref{sec:probform}, we formulate the problem and propose a method for solving it. In Section \ref{sec:B1}, we present the case where the system has a per-link buffer of size one, and provide analytical and numerical results. In Section \ref{sec:B2}, we present numerical results for the case where the system has a per-link buffer of size two and observe similar behavior to the buffer size one case. For the $B=2$ scenario, we also provide a partial analysis. In Section \ref{sec:Decoherence}, we introduce a simple technique for modeling quantum state decoherence and associated cut-off times on qubit storage, and use 
it to examine the effect of decoherence on the bipartite-tripartite capacity region for systems with per-link buffer sizes one and two.
For the former, we also have analytical results. We make concluding remarks in Section \ref{sec:conclusion}.

%% file: BackgroundRelatedWork.tex
Bell states are an integral part of a diverse set of distributed quantum applications, including Quantum Key Distribution (QKD) \cite{bennett1992quantum,ekert1991quantum}, superdense coding \cite{bennett1992communication}, teleportation \cite{bennett1993teleporting}, and distributed quantum computation \cite{jiang2007distributed}. Similarly, GHZ states can be used to implement a variety of quantum protocols, such as cryptographic conferencing \cite{grasselli2018finite}, quantum sensing \cite{eldredge2018optimal}, and multipartite generalization of superdense coding \cite{hao2001controlled}. The advantage of these applications is that they offer functionality that cannot be achieved classically, \emph{e.g.}, information-theoretic security. However, quantum distributed tasks typically require reliable transport of quantum states; this can be a significant challenge due to the exponential rate-versus-distance decay \cite{pirandola2017fundamental}. Quantum repeaters positioned between communicating parties alleviate this issue \cite{Guha2015-qo}. In this work, we use the term ``quantum switch'' instead of ``repeater'' to indicate that the former is equipped with entanglement switching logic.

A mathematical model for a quantum switch was originally introduced in \cite{vardoyan2019stochastic}. There, the authors study a switch that serves only BSMs, but they account for quantum state decoherence, link heterogeneity, and arbitrary buffer sizes (including infinite). In \cite{nain2020multipartite}, the authors study a multipartite entanglement distribution switch that serves $n$-partite GHZ states to users, for $n\geq3$. In this work, links are assumed to be identical and the effects of state decoherence negligible. In both \cite{vardoyan2019stochastic} and \cite{nain2020multipartite}, the authors model the switch as a continuous-time Markov chain (CTMC). In \cite{vardoyan2020exact}, the authors use a discrete-time Markov chain (DTMC) instead, but focus on the simplest and most idealized scenario of bipartite entanglement switching with an infinite number of quantum memories, unit quantum state fidelities, and identical links. In this work, the authors discover that the DTMC is a logically more accurate model of a quantum switch; at the same time, the authors illustrate the modeling and analytical challenges that arise with using a DTMC, rendering more complex problem formulations (\eg, those that include heterogeneous links, finite buffers, and storage cut-offs) intractable.

In contrast to this prior work, we no longer assume that the switch serves only one type of entangled state, \emph{i.e.}, we allow $n$ to be \emph{either} two \emph{or} three, and our goal is to design and evaluate a suitable switching policy.
Another difference is that the quantum switch is assumed to have an infinite number of quantum memories in \cite{nain2020multipartite}, while we consider finite buffer sizes that scale with the number of links.
We opt for continuous-time Markov chains as the modeling technique, because of the findings of \cite{vardoyan2020exact}. It is worth noting that results of the CTMC analyses of \cite{vardoyan2019stochastic} and \cite{nain2020multipartite} have been partially validated using NetSquid, a discrete-event simulation framework for quantum networks \cite{coopmans2020netsquid}. In particular, the authors of \cite{coopmans2020netsquid} showed that CTMCs yield accurate results for the capacity of a quantum switch when there is no quantum state decoherence  (storage cut-off times were not simulated).

Decoherence-associated quantum storage cut-off times \cite{collins2007multiplexed,kozlowski2020designing,khatri2019practical,li2020efficient,rozpkedek2018parameter,rozpkedek2019near} play an important role when it comes to the fidelity (``quality'') of quantum states. The qubit storage cut-off time has varying definitions in the literature: while some define it as the quantum memory lifetime (or coherence time), others view it as a configurable parameter that specifies the amount of time a qubit is to be held in memory; in either case, when the storage cut-off time expires, the qubit is deemed unusable and is discarded. We model this (deterministic) qubit discarding procedure by a probabilistic one, by assuming that the qubit is discarded after an exponentially-distributed amount of time with mean $1/\alpha$. This approach was previously explored in \cite{9351761}, where the authors argued that such an approximation is reasonable for realistic use cases (\ie, in scenarios where the rate at which qubits are discarded is at least an order of magnitude smaller than the entanglement generation rate). In this work, we do not study the effects of decoherence and of BSM and GHZ\footnote{In general, performing a measurement on two or more qubits results in a state with a lower fidelity.} measurements on the quality of the resulting quantum states and focus only on the effects of storage cut-offs on the capacity region of the switch.

%% file: ProbForm_extend.tex
We consider a switch that connects $k$ users over $k$ separate, identical links. Creation of end-to-end entanglement requires two steps. First, two-qubit Bell states are generated pairwise between a qubit stored locally at the switch and a qubit owned by a user. This can be accomplished using a number of available methods, see, \eg, \cite{munro2015inside} and references therein for an overview. Once such link-level two-qubit entangled states have been created, the switch performs joint (entangling) measurements (over $j \geq 2$ locally-held qubits that are entangled with qubits held by $j$ distinct users), which, if successful, produces a $j$-qubit maximally-entangled state between the corresponding $j$ users. Link-level entanglement generation, as well as entangling measurements, when realized with practical systems, are inherently probabilistic~\cite{Guha2015-qo}. We assume that only two-user (two-qubit) and three-user (three-qubit) entangled states are created, \ie, BSMs and 3-qubit GHZ basis measurements are done at the switch\footnote{In principle, by including non-Clifford quantum logic at network nodes, it is possible to extend the entanglement distribution protocol to include a larger class of (non-stabilizer) states. This extension is left as a subject for future work.}. For simplicity, we assume that these $j = 2$ or $3$ qubit measurements at the switch take negligible time and always succeed. The reasoning for the former assumption is that entanglement generation with remote nodes is likely to be more time consuming than local quantum gates and measurements performed at the central switch node; see, \eg, \cite{dahlberg2019link} and \cite{coopmans2020netsquid} for detailed descriptions of timings.
The purpose of the latter assumption is to reduce clutter during the analysis; it
may be relaxed if we allow BSMs and 3-qubit GHZ basis measurements to succeed with 
probabilities $q_1$ and $q_2$,
as the only consequence would be that the bipartite and tripartite capacities $C_2$ and $C_3$ would be scaled by
their respective factors.

Each link attempts two-qubit entanglement in each time slot of length $\tau$ seconds, and with probability $p$, establishes one entangled pair successfully. For simplicity, we model the time to successfully create a link entanglement as an exponential random variable with mean $1/\mu = \tau/p$. Because all links are identical, they all have the same parameters $p$ and $\mu$.
Once a link-level entanglement is successfully generated, the qubits are stored in memories, one at the user, the second at the switch. 
We assume that each link can store $B=1$ or $B=2$ qubits. We also assume that qubits at the switch can decohere and be discarded, and model decoherence (or quantum storage cut-off) time as an exponential r.v. with mean $1/\alpha$. 
Last, when a qubit is stored at the switch, with its entangled pair stored at a user, we refer to this as a {\em stored link entanglement}.  


We assume all possible bipartite and tripartite user entanglement is of interest and consider two classes of probabilistic policies, one for $B=1$ and the second for $B=2$, that provide the flexibility to generate both types of entanglement with arbitrary rates. Policies in both classes incorporate the {\em oldest link entanglement first} (OLEF) rule whereby when a link entanglement is created it is always matched up with stored link entanglement when possible rather than be stored. This has the nice consequence, when coupled with the assumption that links are homogeneous but statistically independent, that the system can be modeled by a continuous time Markov chain where the state simply tracks the number of stored EPR pairs for two users. The next section describes the class of policies for $B=1$ and Section \ref{sec:B2} for the class of $B=2$ policies.

%% file: B1Descr_extend.tex
In this section, we assume that each link can store one qubit in the buffer, so that the per-link buffer size $B = 1$. We model this system using a CTMC, and by obtaining its stationary distribution, we are able to compute the 
capacity region of the switch.
We discover that it is always possible to configure a randomized policy that outperforms TDM, although as the number of links grows, the advantage of using such a policy diminishes.

\subsection{Description}
\begin{figure}
\centering
\includegraphics[width=0.6\textwidth,]{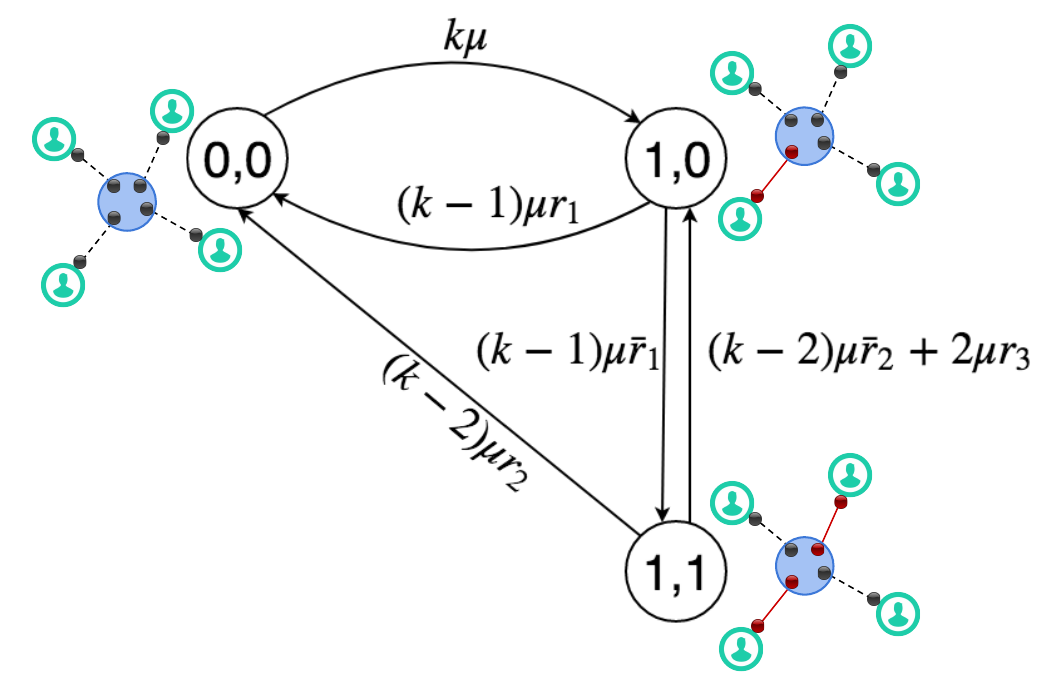}
\caption{CTMC for a system with at least three links and buffer size one for each link. $k$ is the number of links, $\mu$ is the rate of entanglement generation, and $r_1$, $r_2$, and $r_3$ are parameters that specify the scheduling policy.}
\label{fig:homog2dMCB1}
\end{figure}
In a system where the switch can make tripartite measurements, we track two variables for each state of the CTMC: each representing a link with a stored qubit. Hence, $(1,1)$ represents the state where two of the $k$ links have a qubit stored, one each. Note that we do not need to keep track of all links individually due to the OLEF rule and link homogeneity assumption. States $(1,0)$ and $(0,0)$ represent cases where only one link has a stored qubit or no link has a qubit, respectively. 

The system is fully described in Figure \ref{fig:homog2dMCB1}. For a variable $x\in[0,1]$, we use the notation $\bar{x}\equiv 1-x$. When the system is in state $(0,0)$, new entanglement is generated with rate $k\mu$; this is the transition rate from $(0,0)$ to $(1,0)$. When the system is in state $(1,0)$, any new entanglement generated on the link that already has one stored qubit causes the switch to drop one of the qubits. New entanglement on other links is generated with rate $(k-1)\mu$, and the switch must decide whether to immediately use the two qubits for a BSM or keep both and wait for a new link entanglement. To generalize the policy as much as possible, we add a policy parameter, $r_1\in[0,1]$, that specifies the fraction of time the switch performs a BSM. Note that $r_1=1$ corresponds to the policy of always using qubits for BSMs. While this maximizes the bipartite capacity $C_2$, it also means that the tripartite capacity $C_3$ is equal to zero.

Now, suppose that the system is in state $(1,1)$ and a third link generates an entanglement. This event occurs with rate $(k-2)\mu$. The switch has two choices: either use all three qubits for a tripartite measurement, or choose two of them for a BSM. We add another policy parameter, $r_2\in[0,1]$, that specifies fractions of times the switch performs a BSM and three-qubit GHZ measurements in the event of three qubits on three different links. Another event that can occur in the $(1,1)$ state is the generation of entanglement on either of the two links that already have stored entanglement. This event occurs with rate $2\mu$. Since $B=1$, the switch cannot store the new entanglement. A decision must be made: to either discard one of the link entanglements (and remain in state $(1,1)$) or perform a BSM on two of them and keep the third (and transition to state $(1,0)$). Since it is not clear which policy is most advantageous, we add another parameter, $r_3\in[0,1]$, which specifies the fraction of time that the switch performs a BSM when it resides in this state.

%% file: NumericalB1_extend.tex
\begin{figure}
 \centering
 \includegraphics[width=0.6\textwidth]{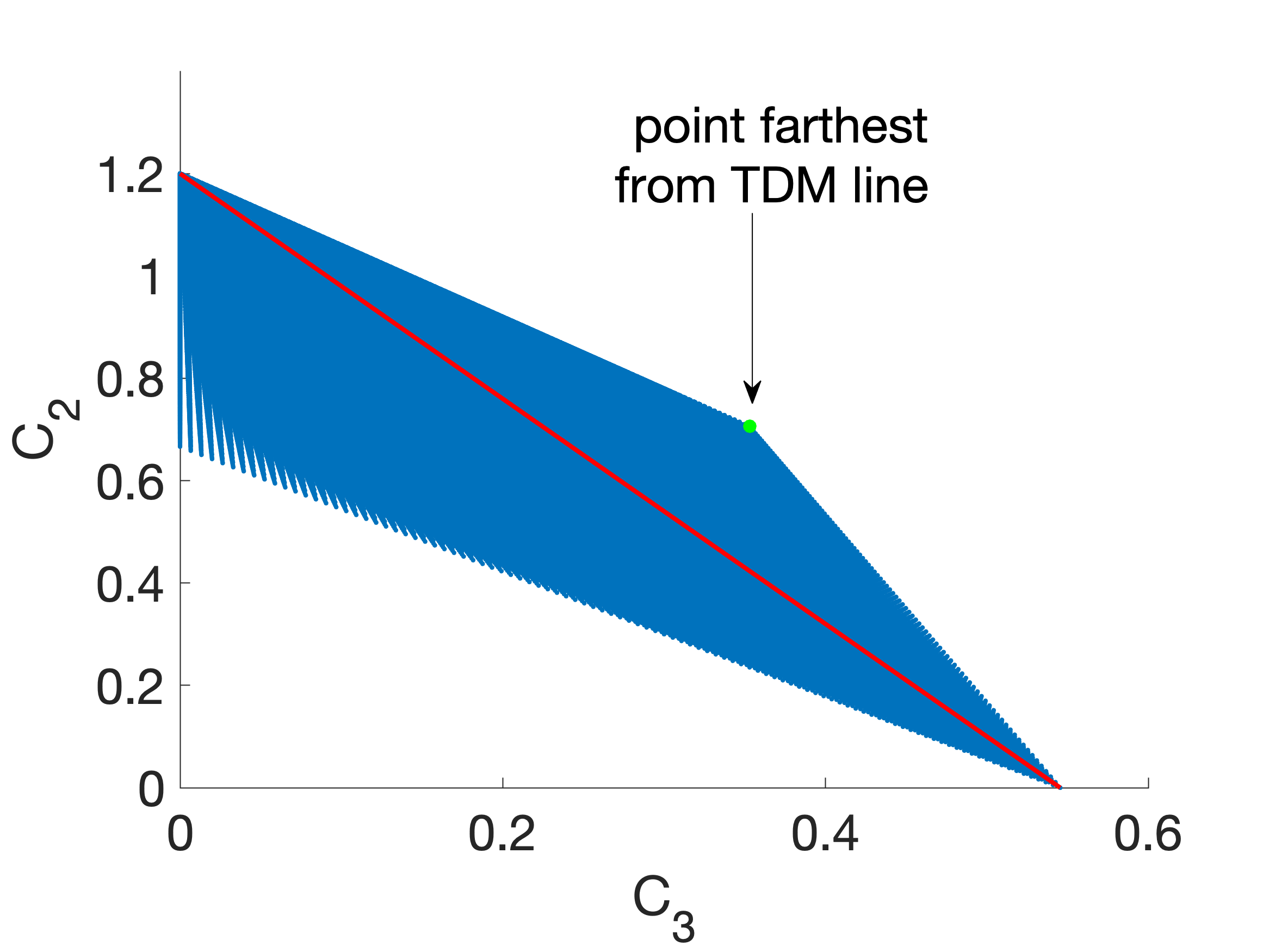}
 \caption{Capacity region for a system of buffer size one and three links. The red line represents the set of TDM policies. $C_2$ and $C_3$ are bipartite and tripartite capacities, respectively.}
    \label{fig:B1plot1}
 \end{figure}
We plot the capacity region for the switch with $B=1$ for all values of $r_1$, $r_2$, $r_3\in[0,1]$ and compare it against TDM. The entanglement generation rate $\mu$ simply scales the capacities, so we set it equal to one (\eg, Kilo-ebits/sec). In Figure \ref{fig:B1plot1}, the number of links is three, and the TDM line is shown in red. Clearly, it is possible to design a policy that yields better performance than TDM: the triangular blue region above the TDM line represents the maximum capacities of the set of such policies.
 
Note that TDM connects points $(0,C_2^*)$ and $(C_3^*,0)$, where $C_2^*$ and $C_3^*$ are the maximum achievable capacities for bipartite and tripartite entanglement, respectively. The point farthest from and above the TDM line (the vertex of the triangular region above the line, shown in green in Figure \ref{fig:B1plot1}) is achieved by setting $r_1=0$ and $r_2=r_3=1$. In other words, the most ``efficient'' policy in terms of being the farthest from the TDM line is the following: $(i)$ never perform BSMs in state $(1,0)$; and $(ii)$ when in state $(1,1)$ and a third entanglement is generated on a \emph{different} link, always use it in a tripartite measurement, but when a third entanglement is generated on one of the links that already has a stored qubit, \emph{always} perform a BSM. Note that the latter rule directs the switch to not waste an entanglement whenever it is possible to use it in a measurement.
\if{false}
 \begin{figure}[htbp]
 \centering
\includegraphics[width=0.35\textwidth]{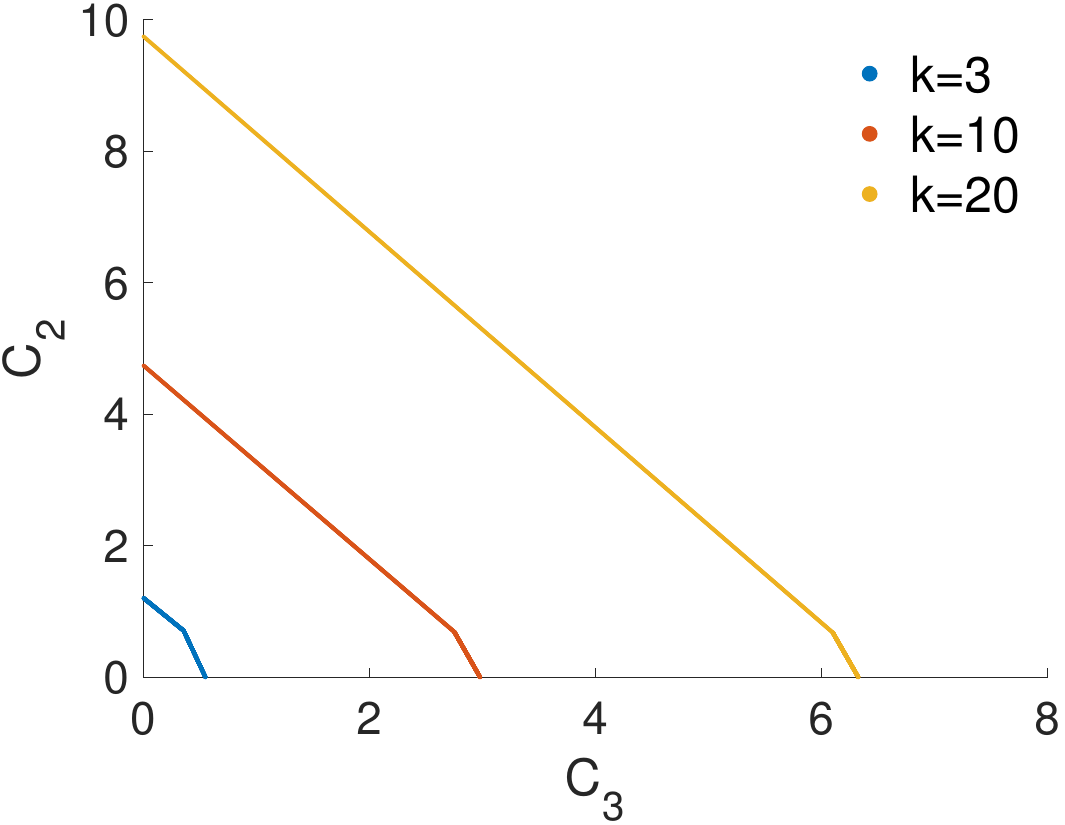}
 \caption{Capacity region for a system of buffer size one and varying number of links. Only the upper-bounding curves are shown.}
    \label{fig:B1plot2}
\end{figure}
 \fi
 \begin{figure}[t]
 \centering
 \subfloat[][$k=10$]{
\includegraphics[width=0.49\textwidth]{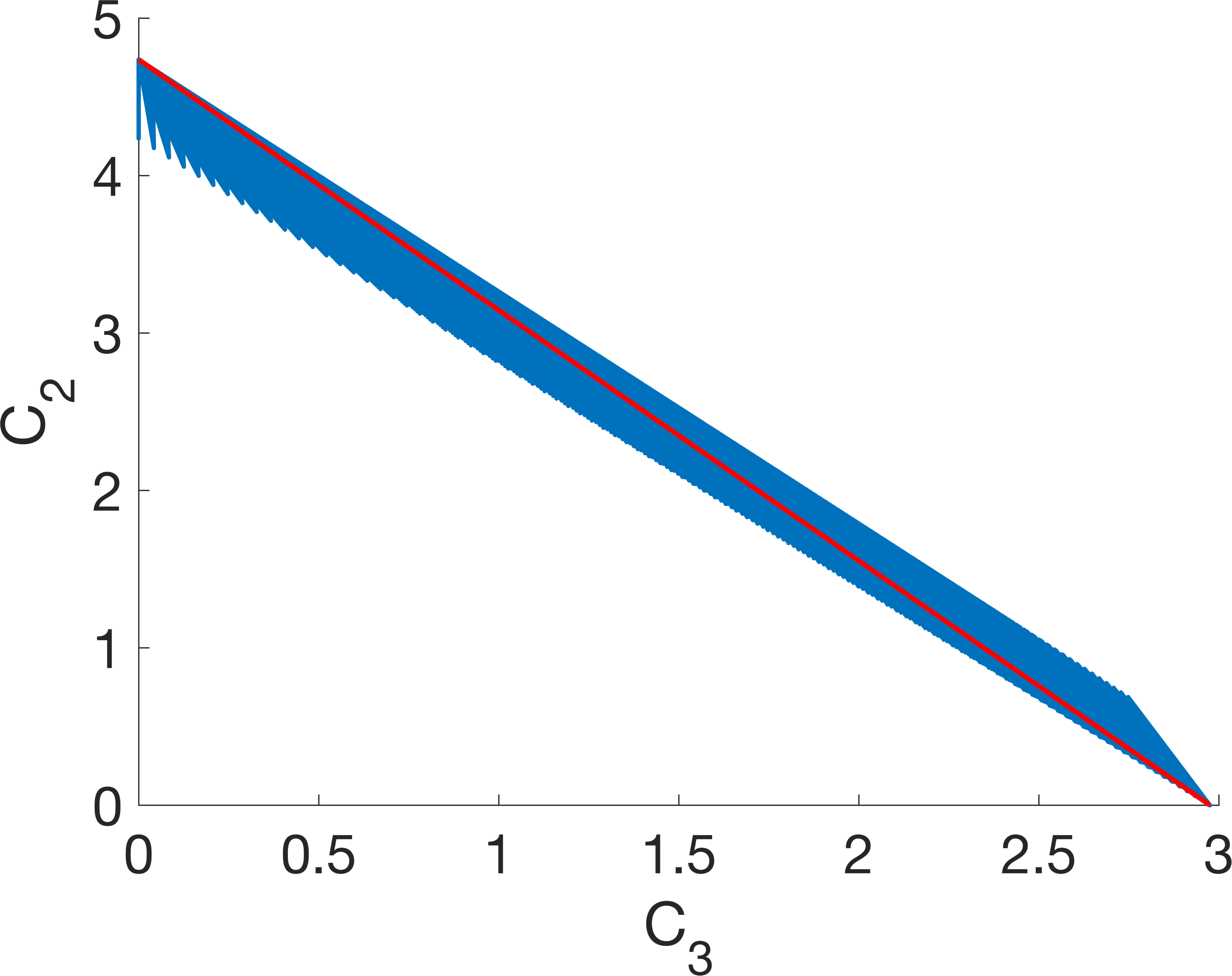}
 }
 \subfloat[][$k=50$]{
\includegraphics[width=0.49\textwidth]{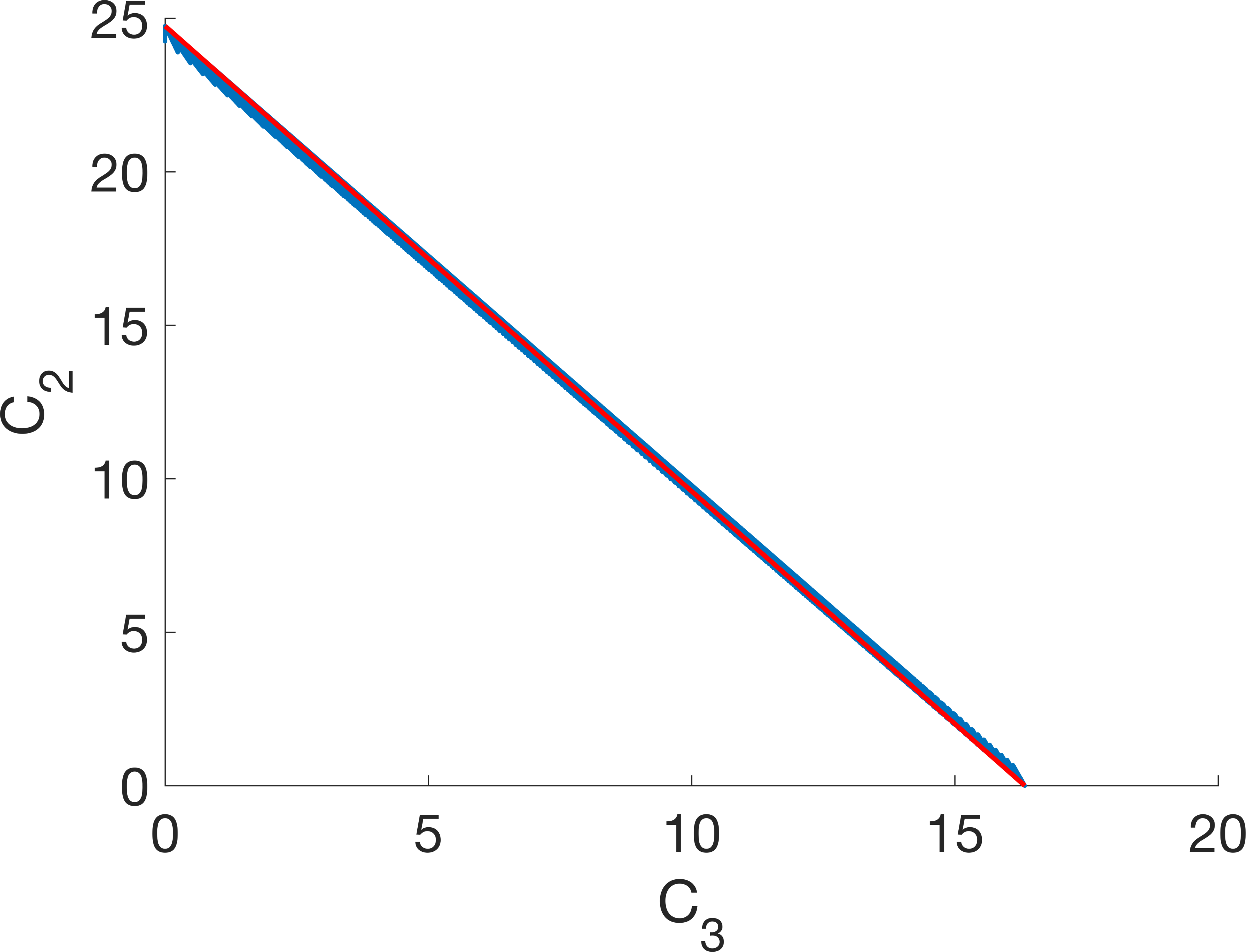}
 }
 \caption{Capacity region for a system of buffer size one and varying number of links. The red line represents the set of TDM policies. The vertex of each triangular region above the TDM line yields the optimal policy in the sense of being farthest from the TDM line.}
   \label{fig:B1plot2}
\end{figure}

The capacity regions for $k=10$ and $50$ are shown in Figure \ref{fig:B1plot2}. Note that as the number of links increases, the differences between TDM and the more efficient random policies diminish. In the next section, we provide an analytical proof of this phenomenon.

%% file: Analysis_extend.tex
Let $\pi(0,0)$, $\pi(1,0)$, and $\pi(1,1)$ represent the stationary distribution of the CTMC in Figure \ref{fig:homog2dMCB1}.
The balance equations (excluding $\mu$, as it cancels out due to every transition rate being its multiple), are:
\begin{align*}
&\pi(0,0)k = \pi(1,0)(k-1)r_1+\pi(1,1)(k-2)r_2,\\
&\pi(1,1)((k-2)r_2+(k-2)\bar{r}_2+2r_3) = \pi(1,0)(k-1)\bar{r}_1,\\
&\pi(0,0)+\pi(1,0)+\pi(1,1) = 1.
\end{align*}
Solving these equations yields
\begin{align*}
&\pi(1,1) = \frac{k(k-1)\bar{r}_1}{D_1},\\\\
&\pi(1,0) = \frac{k(k-2+2r_3)}{D_1},\quad\text{where}\\\\
&D_1 = (k-2+2r_3)((k-1)r_1+k)+(k-1)\bar{r}_1((k-2)r_2+k).
\end{align*}
Then the bipartite and tripartite capacities for this system, $C_2 \equiv C_2(r_1,r_2,r_3)$ and $C_3\equiv C_3(r_1,r_2,r_3)$, are
\begin{align*}
C_2 &= \pi(1,0)(k-1)\mu r_1 + \pi(1,1)((k-2)\mu\bar{r}_2+2\mu r_3)\\\\
&=\frac{k(k-1)\mu (k-2+2 r_3-(k-2)r_2\bar{r}_1)}{(k-2+2r_3)((k-1)r_1+k)+(k-1)\bar{r}_1((k-2)r_2+k)},\\\\
C_3 &= \pi(1,1)(k-2)\mu r_2\\\\
&=\frac{k(k-1)(k-2)\mu r_2\bar{r}_1}{(k-2+2r_3)((k-1)r_1+k)+(k-1)\bar{r}_1((k-2)r_2+k)}.
\end{align*}

\begin{claim}
The maximum value of $C_2$ is given by $C_2^* = C_2(r_1,0,1)=C_2(1,0,r_3)$, where $r_1$ and $r_3$ are arbitrary values in $[0,1]$. The maximum value of $C_3$ is given by $C_3^*=C_3(0,1,0)$.
\end{claim}
\begin{proof}
We start by proving this for $C_2$. First, note that to maximize $C_2$'s numerator and minimize its denominator, $r_2$ must be set to 0. This yields
\begin{align*}
C_2(r_1,0,r_3) &= \frac{k(k-1)\mu (k-2+2 r_3)}{(k-2+2r_3)((k-1)r_1+k)+k(k-1)\bar{r}_1}\\\\
&= \frac{k(k-1)\mu}{(k-1)r_1+k+\frac{k(k-1)\bar{r}_1}{k-2+2r_3}}.
\end{align*}
Now, $r_3=1$ maximizes $C_2(r_1,0,r_3)$, which yields
\begin{align*}
C_2(r_1,0,1) &= \frac{k(k-1)\mu}{(k-1)r_1+k+(k-1)\bar{r}_1} = \frac{k(k-1)\mu}{2k-1}.
\end{align*}
Note that $C_2(1,0,r_3)$ yields the same expression as $C_2(r_1,0,1)$. Next, consider the expression for $C_3$. To minimize the denominator, we should set $r_3=0$. This yields
\begin{align*}
C_3(r_1,r_2,0) &= \frac{k(k-1)(k-2)\mu r_2\bar{r}_1}{(k-2)((k-1)r_1+k)+(k-1)\bar{r}_1((k-2)r_2+k)}\\\\
&= \frac{k(k-1)(k-2)\mu r_2\bar{r}_1}{(k-1)r_1((k-2)\bar{r}_2-k)+k(k-2)+(k-1)((k-2)r_2+k)}.
\end{align*}
It is clear that $r_1$ must be 0, which yields
\begin{align*}
C_3(0,r_2,0) &= \frac{k(k-1)(k-2)\mu r_2}{k(k-2)+(k-1)((k-2)r_2+k)}\\\\
&= \frac{k(k-1)(k-2)\mu r_2}{k(2k-3)+(k-1)(k-2)r_2}\\\\
&= \frac{k(k-1)(k-2)\mu}{\frac{k(2k-3)}{r_2}+(k-1)(k-2)}.
\end{align*}
From above, we can see that $r_2$ must be 1, so the maximum is at $C_3^*=C_3(0,1,0)$. 
\end{proof}

For brevity, let $(C_3(0,1,1),C_2(0,1,1))\equiv (\hat{C}_3,\hat{C}_2)$; this is the point farthest above the TDM line within the achievable capacity region (\emph{e.g.}, the green point in Figure \ref{fig:B1plot1}). We prove this as part of the proof of the claim below.

\begin{claim}
Any point $(C_3,C_2)$ in the achievable capacity region satisfies the following constraints:
\begin{align}
C_2 &\leq -\frac{3k-2}{2k-1}C_3+\frac{\mu k (k-1)}{2k-1} \quad\text{and}\label{eq:bound1B1}\\\nonumber\\
C_2 &\leq -\frac{k(k-2)+2(k-1)^2}{k(k-2)}C_3 + \mu (k-1),\label{eq:bound2B1}\\\nonumber\\
C_2,~C_3&\geq 0.
\end{align}
Moreover (\ref{eq:bound1B1}) and (\ref{eq:bound2B1}) define a tight upper bound on the achievable capacity region.
\end{claim}
\begin{proof}
First, we must show that the point $(\hat{C}_3,\hat{C}_2)$ is indeed the farthest point from the TDM line.
 We do this by first assuming that there exists an achievable capacity region above the TDM line that is shaped like a triangle, as in Figures \ref{fig:B1plot1} and \ref{fig:B1plot2}, and later prove this to be true. Moreover, the sides of this triangular region impose upper bounds on the achievable bipartite-tripartite capacities.
With these assumptions proven, it is clear that the point farthest from the TDM line will be located at the vertex of this triangle.
Thus, let us find a point $(C_3,C_2)$ on the plane such that the (potentially negative) slope of the line that passes through it and $(0,C_2^*)$ is maximized. This is equivalent to minimizing the quantity
\begin{align*}
\frac{C_2^*-C_2}{C_3} &= \frac{(3k-2)(k-2)r_2+2(k-1)(1-r_3)}{r_2(2k-1)(k-2)}.
\end{align*}
To do so, we must set $r_3=1$, yielding a slope of
\begin{align}
-\frac{C_2^*-C_2}{C_3} &= -\frac{3k-2}{2k-1}.
\label{eq:slopeFar}
\end{align}
Next, note that the TDM line is given by the equation 
\begin{align}
f(x,y)=y-C_2^*(1-x/C_3^*), \label{eqn:tdmline}
\end{align} 
and the distance between it and any point $(C_3,C_2)$ is given by 
\begin{align*}
\frac{\lvert f(C_3,C_2)\rvert}{\sqrt{1+(C_2^*/C_3^*)^2}}.
\end{align*} 
Hence, it is sufficient to maximize $\lvert f(C_3(r_1,r_2,1),C_2(r_1,r_2,1))\rvert$, given by
\begin{align*}
\frac{2\mu k (k-1)}{(2k-1)\left(k-2+\frac{2k^2-k}{(k-1)r_2(1-r_1)}\right)}.
\end{align*}
It is clear that we must set $r_2=1$ and $r_1=0$, yielding $(\hat{C}_3,\hat{C}_2)\equiv (C_3(0,1,1),C_2(0,1,1))$ as the point farthest from the TDM line, as expected. Note that this holds under the assumption that the achievable capacity region has the shape of a triangle, which still remains to be proven.

To do so, consider the line passing through $(0,C_2^*)$ and $(\hat{C}_3,\hat{C}_2)$, whose slope is given by (\ref{eq:slopeFar}):
\begin{align}
y_1 &= -\frac{3k-2}{2k-1}x_1+\frac{\mu k (k-1)}{2k-1},
\label{eq:line1}
\end{align}
and the line passing through $(\hat{C}_3,\hat{C}_2)$ and $(C_3^*,0)$:
\begin{align}
y_2 &= -\frac{k(k-2)+2(k-1)^2}{k(k-2)}x_2 + \mu (k-1).
\label{eq:line2}
\end{align}
It is not hard to show that for any point $(C_3, C_2)$,
(\ref{eq:bound1B1}) and (\ref{eq:bound2B1}) are satisfied.
In other words, all points in the achievable capacity region fall on or below these two lines. To prove that this upper bound is tight, it remains to show that all points on lines (\ref{eq:line1}) and (\ref{eq:line2}) are achievable. To see this, let $r_1=0$ and $r_3=1$:
\begin{align*}
C_2(0,r_2,1) &= \frac{(k-(k-2)r_2)k(k-1)\mu}{k^2+(k-1)(k+(k-2)r_2)},\\\\
C_3(0,r_2,1) &= \frac{(k-2)r_2k(k-1)\mu}{k^2+(k-1)(k+(k-2)r_2)}.
\end{align*}
Note that any point $(C_3(0,r_2,1),C_2(0,r_2,1))$ is \emph{on} line (\ref{eq:line1}), and these two functions are continuous in $r_2\in[0,1]$.
Similarly, letting $r_1=0$ and $r_2=1$, we have
\begin{align*}
C_2(0,1,r_3) &= \frac{2r_3k(k-1)\mu}{k(k-2+2r_3)+2(k-1)^2},\\\\
C_3(0,1,r_3) &= \frac{k(k-1)(k-2)\mu}{k(k-2+2r_3)+2(k-1)^2}.
\end{align*}
Any point $(C_3(0,1,r_3),C_2(0,1,r_3))$ is \emph{on} line (\ref{eq:line2}), and these two functions are continuous in $r_3\in[0,1]$. Using these facts, we conclude that \emph{all} points on (\ref{eq:line1}) and (\ref{eq:line2}) are achievable. 
\end{proof}
\begin{claim}
As $k\to\infty$, the benefit of using an alternate policy (one that lies above TDM) diminishes.
\end{claim}
\begin{proof}
We prove this by showing that as $k\to\infty$, the ratio of the achievable area above the TDM line, which we call $A_{\triangle}$ (because this area has the shape of a triangle) to the total area below the capacity region, which we call $A_T$, goes to zero. For $A_{\triangle}$, the length of the base of the triangle is simply the distance between the points $(0,C_2^*)$ and $(C_3^*,0)$, or $\sqrt{(C_2^*)^2+(C_3^*)^2}$. The height is given by 
\begin{align*}
\frac{\lvert f(\hat{C}_3,\hat{C}_2)\rvert}{\sqrt{1+(C_2^*/C_3^*)^2}}.
\end{align*} 
Then,
\begin{align*}
A_{\triangle} &= \frac{\lvert f(\hat{C}_3,\hat{C}_2)\rvert C_3^*}{2}.
\end{align*}
Then, the area below the TDM line is given by
\begin{align*}
A_{TDM} &= \frac{C_2^*C_3^*}{2},
\end{align*}
so the total area is
\begin{align*}
A_T &= A_{\triangle}+A_{TDM} = 
\frac{\lvert f(\hat{C}_3,\hat{C}_2)\rvert C_3^*+C_2^*C_3^*}{2}.
\end{align*}
Then the ratio of the area above the TDM to the total area is
\begin{align*}
\frac{A_{\triangle}}{A_T} &=\frac{\lvert f(\hat{C}_3,\hat{C}_2)\rvert C_3^*}{\lvert f(\hat{C}_3,\hat{C}_2)\rvert C_3^*+C_2^*C_3^*}
=\frac{\lvert f(\hat{C}_3,\hat{C}_2)\rvert }{\lvert f(\hat{C}_3,\hat{C}_2)\rvert +C_2^*}
=\frac{1}{1 +\frac{C_2^*}{\lvert f(\hat{C}_3,\hat{C}_2)\rvert }}.
\end{align*}
To prove that this ratio goes to zero with $k$, it suffices to show that the second term in the denominator goes to $\infty$. It can be shown that $C_2^*/\lvert f(\hat{C}_3,\hat{C}_2)\rvert $ simplifies to
\begin{align*}
\frac{3k^2 - 4k + 2}{2(k - 1)}~~ \overset{k\to\infty}\longrightarrow~\infty.
\end{align*}
\end{proof}
Using (\ref{eqn:tdmline}), we may describe the set of TDM policies from a more mathematical perspective. Namely, for any point $(C_3,C_2)$ on the TDM line, we may write $C_2=\delta C_2^{*}$, for $\delta\in[0,1]$, which by (\ref{eqn:tdmline}) implies that $C_3=(1-\delta)C_3^{*}$. Thus, $(C_3,C_2)$ is obtained by a policy that performs BSMs a fraction $\delta$ of the time, and 3-GHZs a fraction $1-\delta$ of the time.

%% file: B2Descr_extend.tex
\begin{figure}
\centering
\includegraphics[width=0.6\textwidth,trim={0 1cm 0 0},clip]{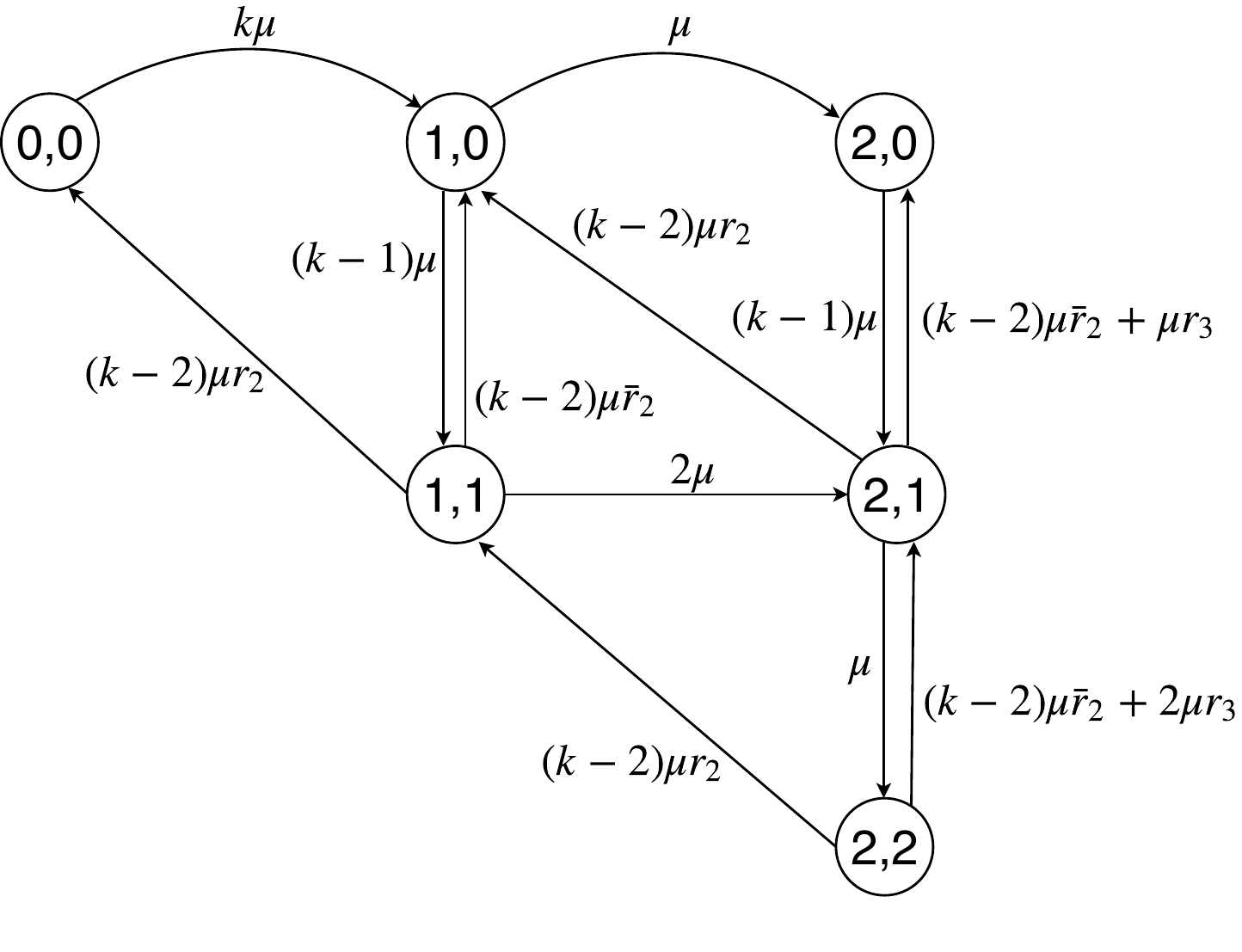}
\caption{CTMC for a system with at least three links and buffer size two for each link. $k$ is the number of links, $\mu$ is the rate of entanglement generation, and $r_2$ and $r_3$ are parameters that specify the scheduling policy.}
\label{fig:homogBuf2}
\end{figure}

In a system with per-link buffer size two, there are three additional states, as shown in Figure \ref{fig:homogBuf2}. Recall that in Section \ref{sec:B1}, we analyzed the Markov chain of buffer size one systems to find policies that are optimal, in the sense that the corresponding $(C_3,C_2)$ point within the capacity region is farthest from the TDM line. The goal of this part of the study, -- \ie, for $B=2$ systems -- is to show the existence of better policies than TDM, rather than to find such an optimal policy, as the former allows for a significantly simpler analysis. Hence, the design in Figure \ref{fig:homogBuf2} does not encapsulate all possible switching policies: for instance, there is no $r_1$ parameter here. 
Our exhaustive policy search over the entire parameter space for systems with $B=1$ in Section \ref{sec:B1} revealed that $r_1$ is best set to zero; while it is possible that a positive $r_1$ may be beneficial to a $B=2$ system, we omit it from the model under the intuitive reasoning that the switch should favor states which allow for (more demanding in terms of the number of necessary EPR pairs) 3-GHZ measurements.
In addition, note that if the system is in state $(1,1)$ and another entanglement is generated on one of the links that already has a stored qubit, the system is not allowed to use two of the qubits for a BSM. The reasoning is that since $B=2$, there is enough space to keep the new qubit. Similarly, when the system is in state $(2,1)$ a BSM is only allowed if $(i)$ another entanglement is generated on one of the $k-2$ links that does not have a stored qubit, or $(ii)$ another entanglement is generated on the link that already has two qubits stored. In the latter scenario, not performing a BSM would cause a qubit to be discarded. While this design does not grant the switch access to the full range of policies, it does enable us to find a class of policies that are more efficient than TDM.

A note on the notation in this and the following sections: with some abuse of notation, we use the variables $C_2$ and $C_3$ to represent the bipartite and tripartite capacities of a switch with per-link buffer size two.
\subsection{Numerical Results}

Figure \ref{fig:B2plots} shows capacity regions for $B=2$ with number of links $k=3$ and $10$. We observe that policies more efficient than TDM can be found, but as the number of links grows, the advantage of such policies relative to TDM diminishes. This phenomenon mimics that of the $B=1$ switch.
\if{false}
\begin{figure}[htbp]
 \centering
 \includegraphics[width=0.35\textwidth]{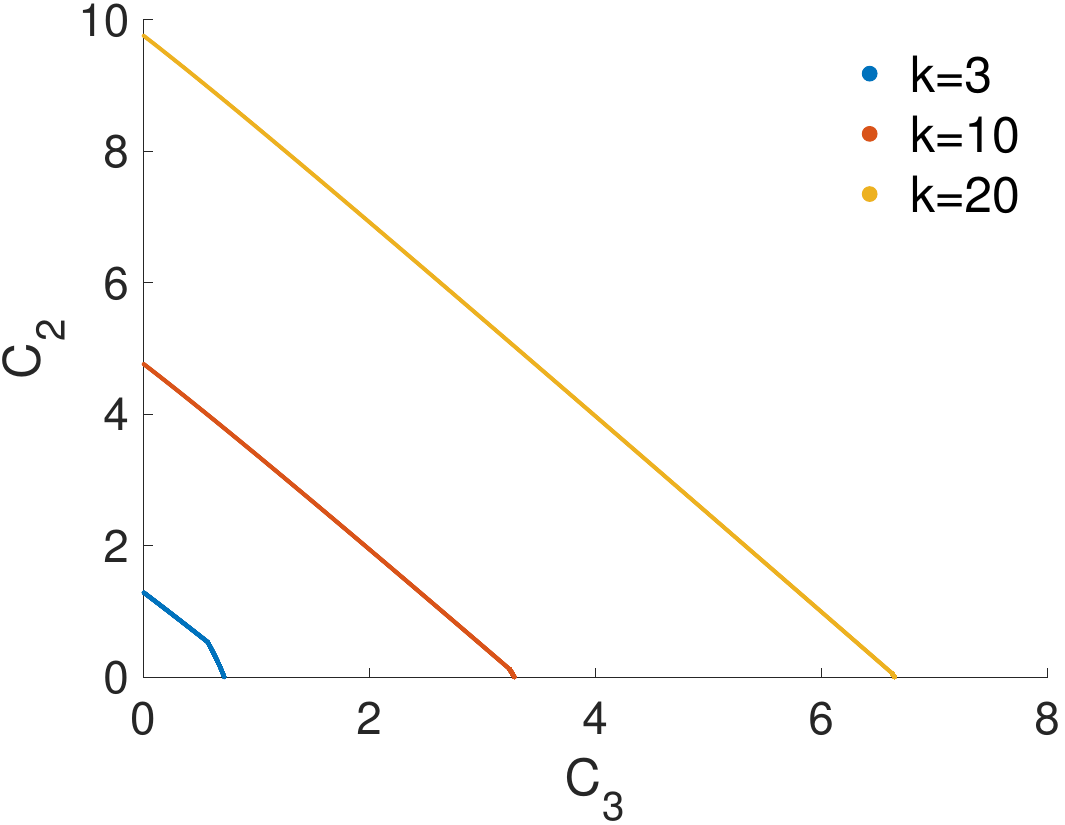}
 \caption{Capacity region for a system of buffer size two and varying number of links. Only the upper-bounding curves are shown.}
   \label{fig:B2plots}
\end{figure}
\fi
\begin{figure}
 \centering
 \subfloat[][$k=3$]{
 \includegraphics[width=0.49\textwidth]{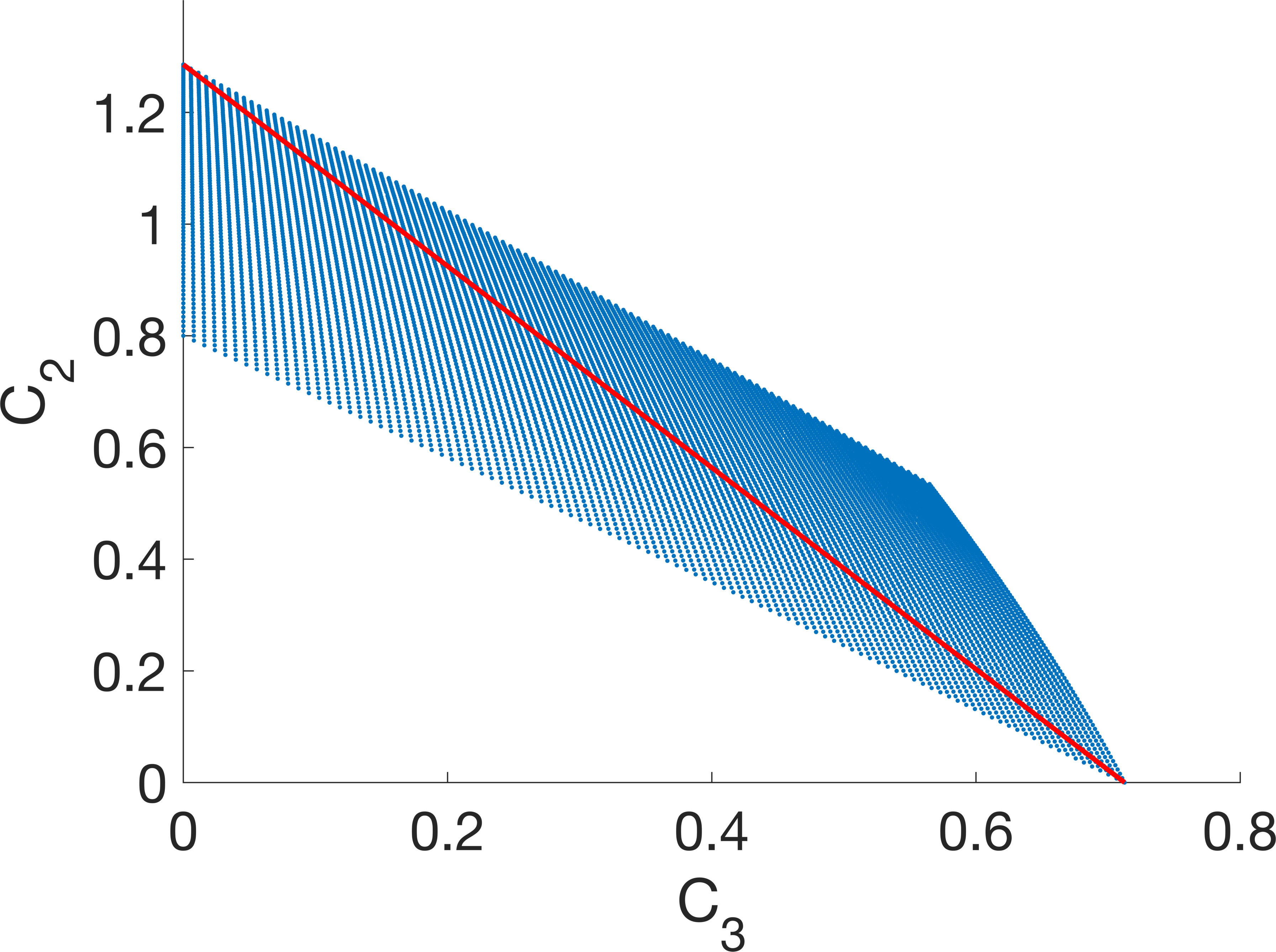}
 }
 \subfloat[][$k=10$]{
\includegraphics[width=0.49\textwidth]{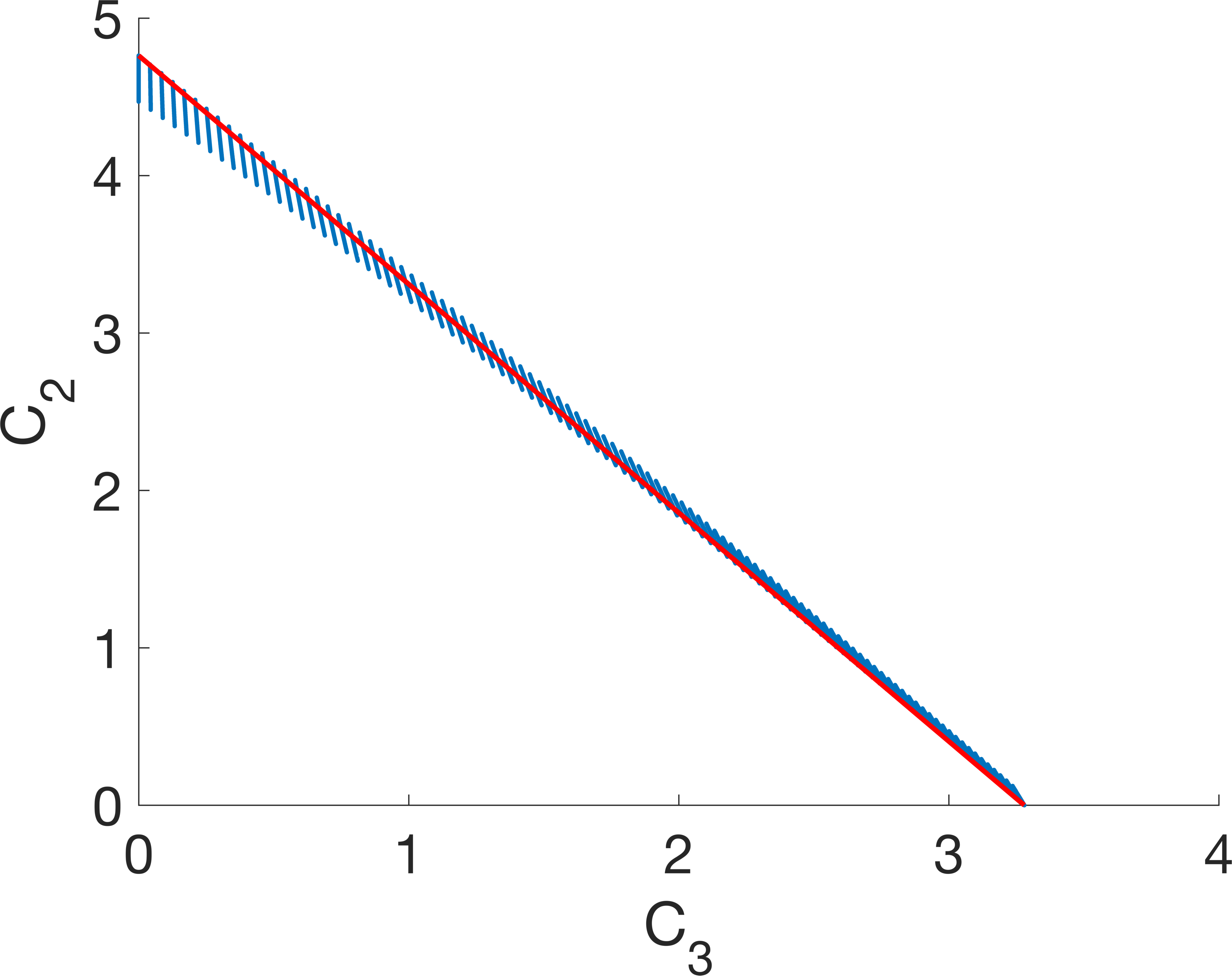}
 }
 \caption{Capacity region for per-link buffer size $B=2$, for $k=3$, $10$ links. The red line represents the set of TDM policies.}
   \label{fig:B2plots}
\end{figure}
\begin{figure}[htbp]
 \centering
 \subfloat[][$k=3$]{
 \includegraphics[width=0.49\textwidth]{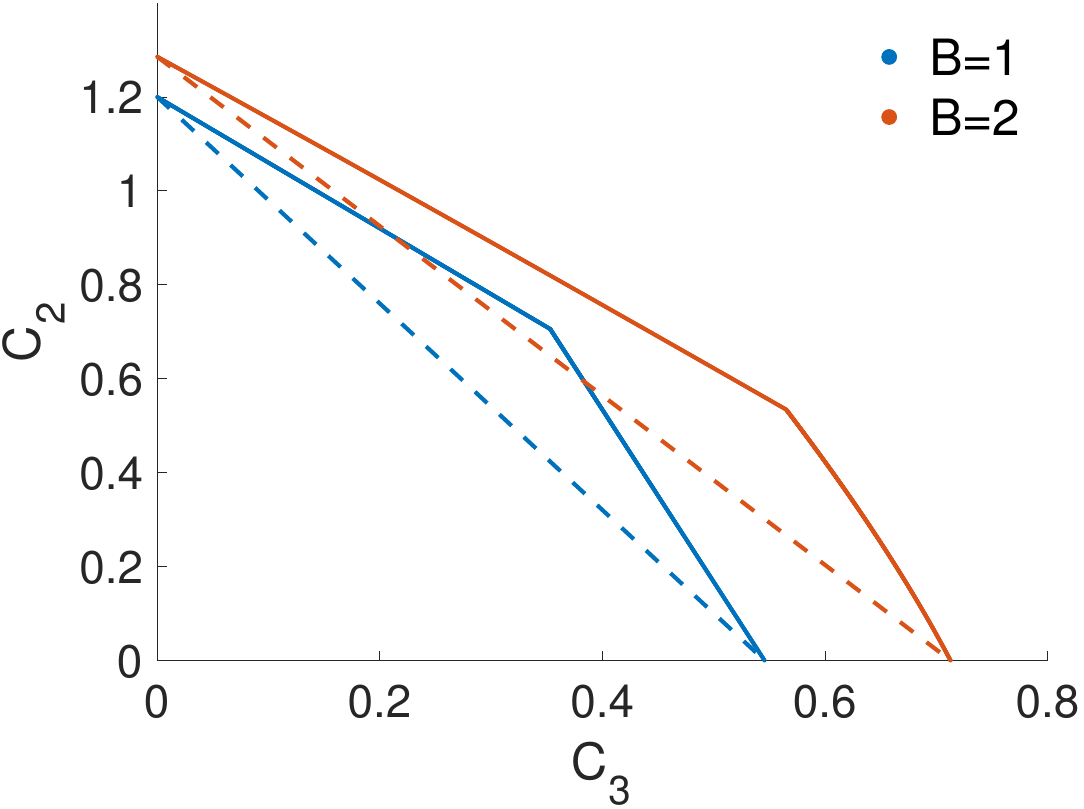}
 }
 \subfloat[][$k=10$]{
\includegraphics[width=0.49\textwidth]{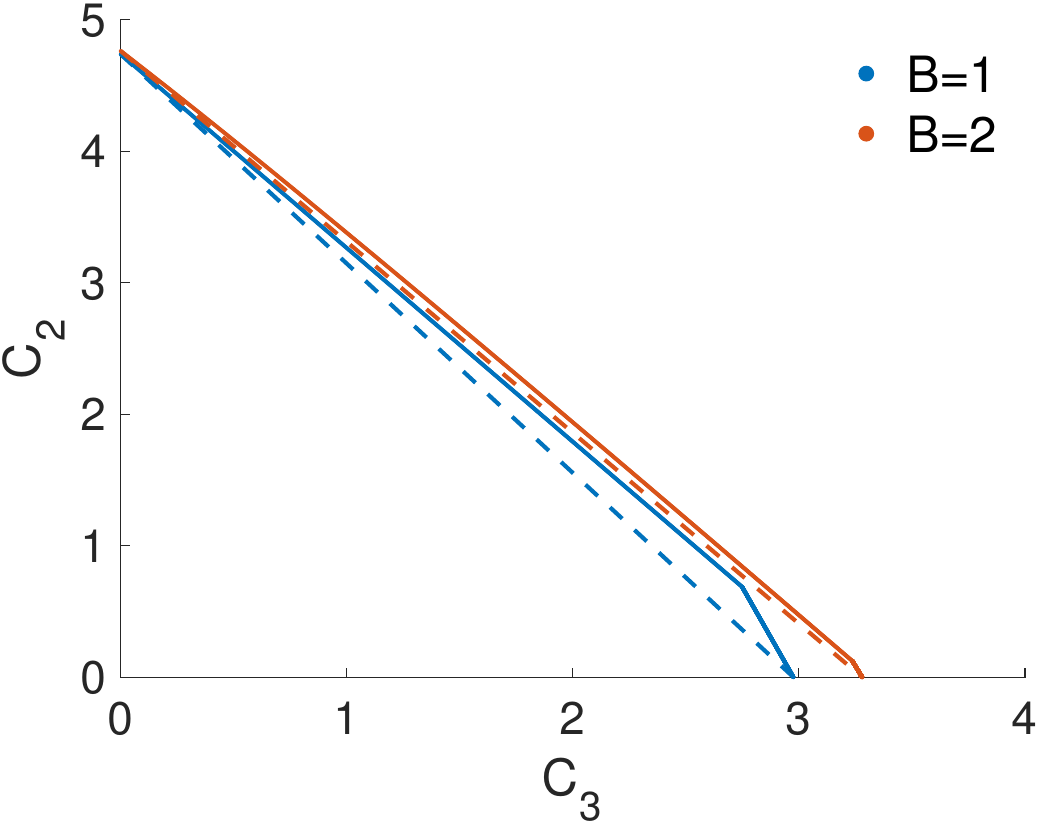}
 }
 \caption{Comparison of capacity regions for systems of buffer sizes one and two with varying number of links $k$, and entanglement generation rate $\mu=1$.}
   \label{fig:B2plotscomp}
\end{figure}
Figure \ref{fig:B2plotscomp} shows a comparison of $B=1$ and $B=2$ switches for three and ten links. We observe that while there is a clear benefit to extra buffer space for a small number of users, the advantage becomes less apparent as the number of users grows. In addition, it appears that $C_3$ benefits more from the extra buffer space than $C_2$.
\subsection{Analysis}
\input{Analysis_B2}

%% file: Analysis_B2.tex
In this section, we provide a partial analysis of the quantum switch with two quantum memories allocated to each link. Specifically, we derive the stationary distribution of the CTMC representing this system, and compute the maximum bipartite and tripartite capacities for the switch. However, we do not fully characterize the capacity region as in the case of $B=1$.

Let $\pi(0,0)$, $\pi(1,0)$, $\pi(2,0)$, $\pi(1,1)$, $\pi(2,1)$, and $\pi(2,2)$ represent the stationary distribution of the CTMC in Figure \ref{fig:homogBuf2}.
The balance equations (excluding $\mu$, as it cancels out due to every transition rate being its multiple), are:
\begin{align*}
&k\pi(0,0) = (k-2)r_2\pi(1,1),\\
& k \pi(1,0) = k\pi(0,0)+(k-2)\bar{r}_2\pi(1,1)+(k-2)r_2\pi(2,1),\\
&(k-1)\pi(2,0) = \pi(1,0)+((k-2)\bar{r}_2+r_3)\pi(2,1),\\
&k\pi(1,1) = (k-1)\pi(1,0)+(k-2)r_2\pi(2,2)\\
&(k-2+2r_3)\pi(2,2) = \pi(2,1)\\
&\pi(0,0)+\pi(1,0)+\pi(2,0)+\pi(1,1)+\pi(2,1)+\pi(2,2)=1.
\end{align*}
Solving these equations yields
\begin{align*}
\pi(0,0) &= \frac{1}{k}\frac{(k-1)(k-2)^2r_2^2((k-1)((k+1)(k-2)+2kr_3)+3k-2)}{D_2},\\
\pi(1,0) &= \frac{k(k-1)(k-2)r_2((k+1)(k-2)+2kr_3)}{D_2},\\
\pi(1,1) &= \frac{(k-1)(k-2)r_2((k-1)((k+1)(k-2)+2kr_3)+3k-2)}{D_2},\\
\pi(2,0) &= \frac{k(k-2)r_2((k+1)(k-2)+2kr_3)+k(k-2+2r_3)(3k-2)((k-2)\bar{r}_2+r_3)}{D_2},\\
\pi(2,1) &= \frac{k(k-1)(3k-2)(k-2+2r_3)}{D_2},\quad\text{and}\\
\pi(2,2) &= \frac{k(k-1)(3k-2)}{D_2},\quad\text{where}\\
D_2 &= kr_2(2kr_3+(k+1)(k-2))(r_2(k-1)(k-2)+2(k-1)^2+k-2)\\
&\quad+(k-2+2r_3)(3k-2)(k\bar{r}_2(2k-3)-r_2^2(k-1)(k-2)+kr_3)+k(k-1)(3k-2).
\end{align*}
The bipartite and tripartite capacities for this system, $C_2 \equiv C_2(r_2,r_3)$ and $C_3\equiv C_3(r_2,r_3)$, are
\begin{align}
C_2 &= \pi(1,1)(k-2)\mu \bar{r}_2+\pi(2,1)((k-2)\mu\bar{r}_2+\mu r_3)+\pi(2,2)((k-2)\mu\bar{r}_2+2\mu r_3)\nonumber\\
&= (k-1)\mu\frac{(k-2)\bar{r}_2[k(3k-2)(k-1+2r_3)+(k-2)r_2((k-1)((k+1)(k-2)+2kr_3)+3k-2)]}{D_2}\nonumber\\
&\quad+(k-1)\mu\frac{(3k-2)k(k+2r_3)r_3}{D_2},\label{eq:C2B2}\\
C_3 &= (k-2)\mu r_2 (\pi(1,1)+\pi(2,1)+\pi(2,2))\nonumber\\
&= (k-1)(k-2)\mu r_2\frac{k(3k-2)(k-1+2r_3)+(k-2)r_2[(k-1)((k+1)(k-2)+2kr_3)+3k-2]}{D_2}\label{eq:C3B2}.
\end{align}
Finally, we note that the
maximum value of $C_2$ is given by $C_2^* = C_2(0,1)$,
and the 
maximum value of $C_3$ is given by $C_3^*=C_3(1,0)$.
Intuitively, setting $r_2=1$ ensures that a BSM is never performed when a 3-qubit GHZ basis measurement could be performed instead, while setting $r_3=1$ preserves all stored qubits for an opportunity of a 3-qubit GHZ measurement in the future. Similarly, setting $r_2=0$ and $r_3=1$ has the opposite effect, causing the switch to take all opportunities to perform a BSM.

%% file: Decoherence_extend.tex
\begin{figure}
   \centering
   \includegraphics[width=0.6\textwidth,trim={0 2cm 0 0},clip]{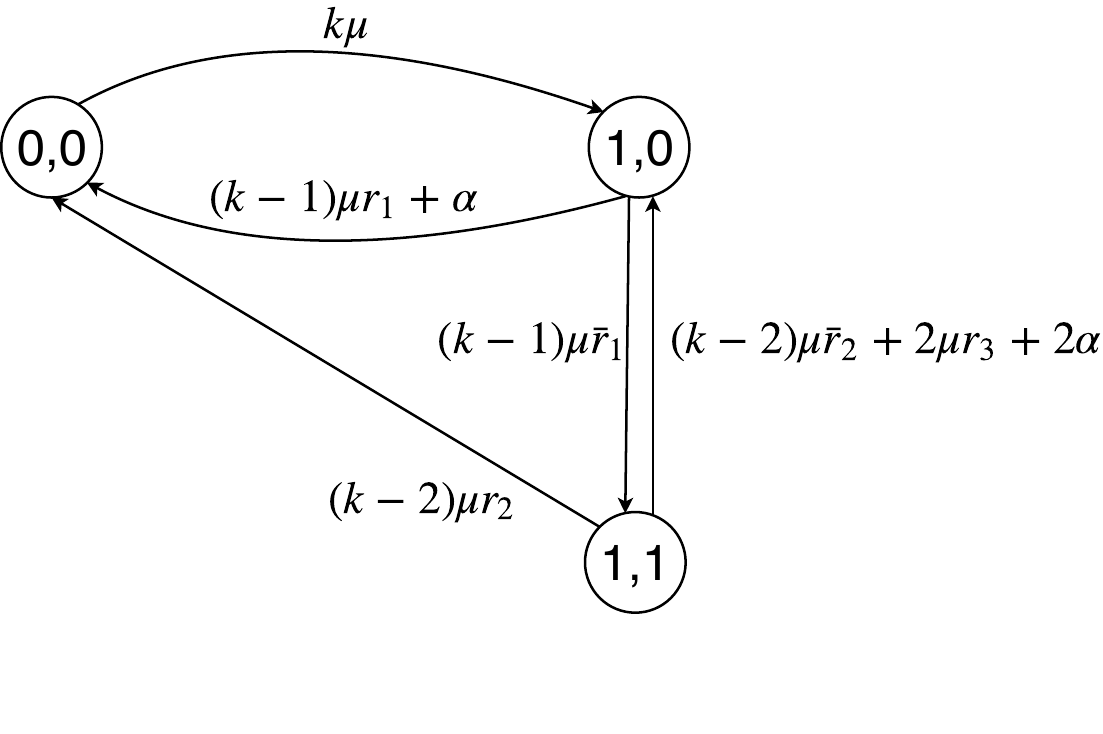}
 \caption{CTMC for a system with at least three links and buffer size one. $k$ is the number of links, $\mu$ is the rate of entanglement generation, $\alpha$ is the decoherence-associated storage cut-off rate, and $r_1$, $r_2$, $r_3$ are parameters that specify the scheduling policy.}
   \label{fig:buf1decoh}
 \end{figure}

In this section, we present a simple way to augment the model from Section \ref{sec:B1} to account for the decoherence of quantum states and associated cut-off time on quantum storage. For switches with $B=1$, we present both analytic and numerical results. 
We also augment the model from Section \ref{sec:B2} to incorporate decoherence and storage cut-offs, but for switches with $B=2$, we present only numerical results.
Our decoherence model is described in Section \ref{sec:probform}. For $B=1$, the resulting CTMC is illustrated in Figure \ref{fig:buf1decoh}.
 
The analysis of this model is almost identical to that of the $B=1$ system without decoherence. As with the latter, the capacity region is bounded above by two lines:
\begin{align*}
y_1 &= -\frac{\mu(3k-2)(\alpha+(k-2)\mu)+2\alpha^2}{\mu(k-2)((2k-1)\mu+\alpha)}x_1+\frac{k(k-1)\mu^2}{(2k-1)\mu+\alpha},\\\\
y_2 &= -\frac{2(k-1)^2\mu^2+(k\mu+\alpha)((k-2)\mu+2\alpha)}{\mu(k-2)(k\mu+\alpha)}x_2+\frac{k(k-1)\mu^2}{k\mu+\alpha}.
\end{align*}
To avoid ambiguity, let $C_2^{\prime}$ and $C_3^{\prime}$ denote the bipartite and tripartite capacities of a system with decoherence.
As with the previous model, $C_2^{\prime}$ is maximized at $r_1=1$, $r_2=r_3=0$; $C_3^{\prime}$ is maximized at $r_1=r_3=0$, $r_2=1$, and the point farthest from TDM is obtained by setting $r_1=0$, $r_2=r_3=1$.
The first bounding line passes through the points $(0,C_2^{\prime}(1,0,0))$ and $(C_3^{\prime}(0,1,1),C_2^{\prime}(0,1,1))$; and the second line passes through $(C_3^{\prime}(0,1,1),C_2^{\prime}(0,1,1))$ and $(C_3^{\prime}(0,1,0),0)$. Moreover, all points \emph{on} the bounding lines are achievable, indicating that the bound is tight.

The capacities are given by
\begin{align*}
C_2^{\prime} &=(k(k-1)\mu^2\left(2(\alpha r_1+\mu r_3)+(k-2)\mu(1-r_2\bar{r}_1)\right))/D,\\
C_3^{\prime} &= (k\mu^3(k-1)(k-2) \bar{r}_1 r_2)/D, \quad\text{where}\\
D&=(k-1)\mu\bar{r}_1((k-2)\mu r_2+k\mu)
+(k\mu+(k-1)\mu r_1+\alpha)((k-2+2r_3)\mu+2\alpha).
\end{align*}
Note that the denominator is quadratic in $\alpha$. This causes both $C_2^{\prime}$ and $C_3^{\prime}$ to tend to zero as $\alpha\to\infty$.
 
 \begin{figure}
 \centering
 \subfloat[][$k=3$]{
 \includegraphics[width=0.49\textwidth]{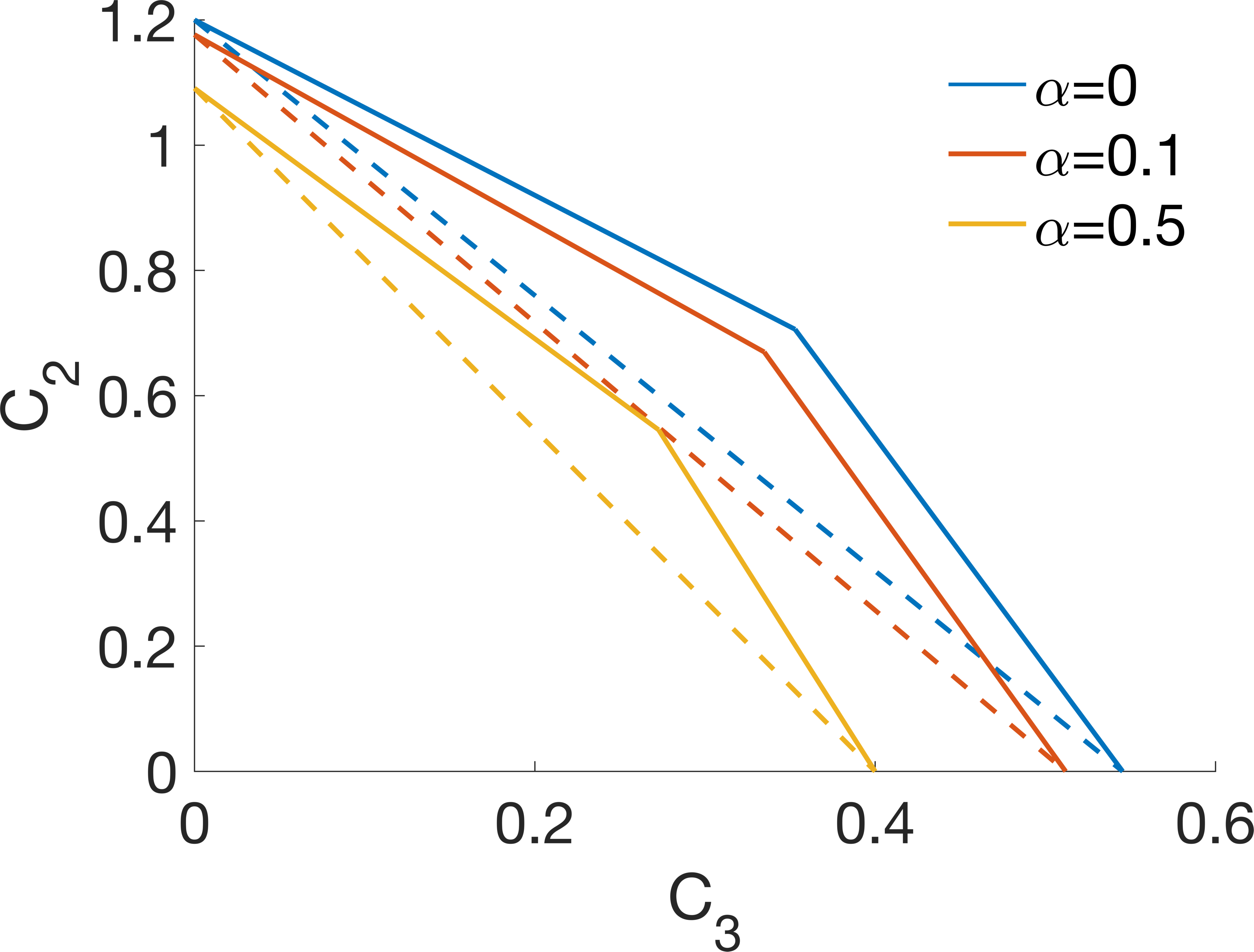}
 }
 \subfloat[][$k=10$]{
\includegraphics[width=0.49\textwidth]{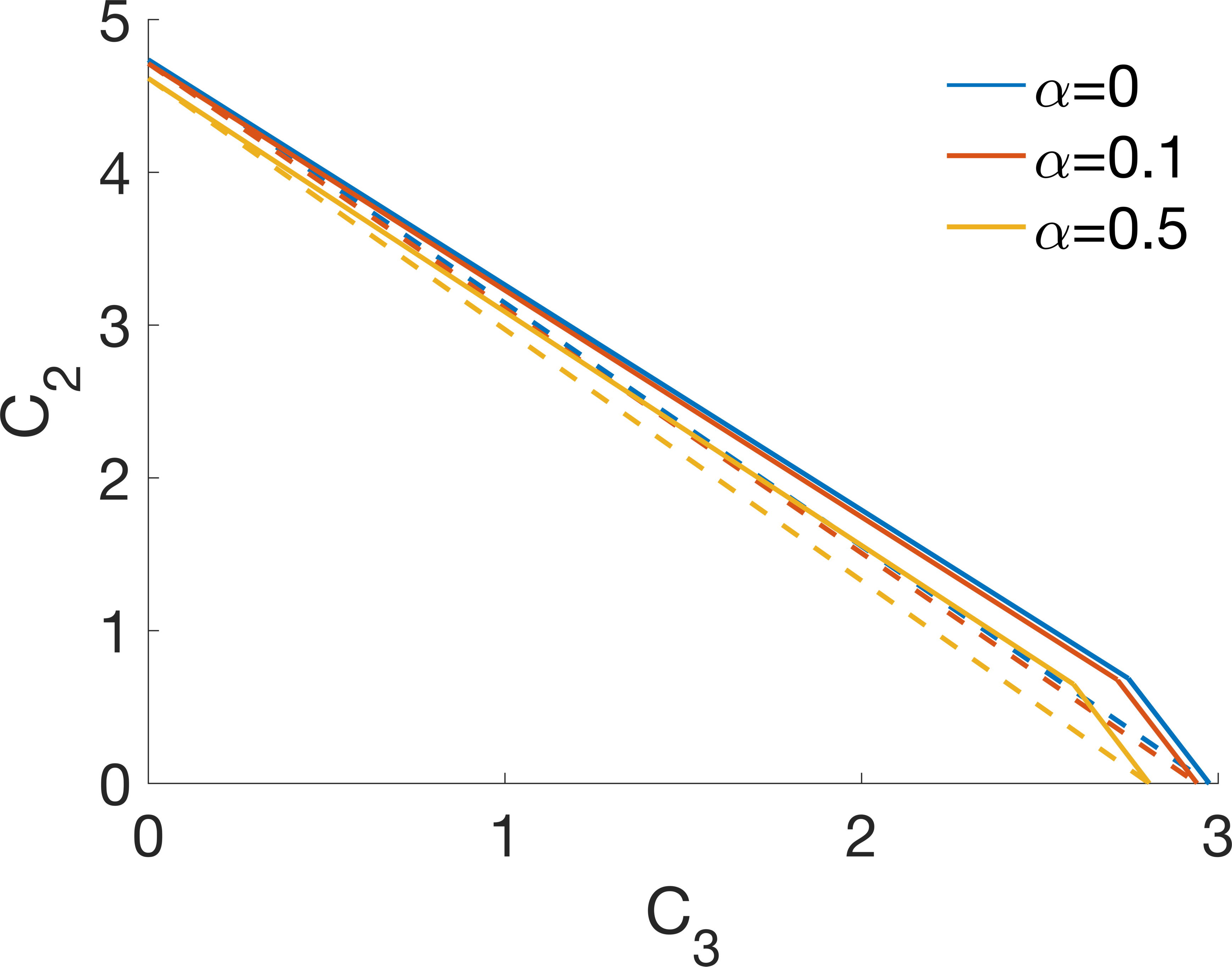}
 }
 \caption{Capacity region for a system of buffer size one and varying number of links $k$, decoherence/storage cut-off rates $\alpha$, and entanglement generation rate $\mu=1$. The solid lines are the upper boundaries of the capacity region, and the dashed are TDM lines.}
   \label{fig:decohPlotsB1}
\end{figure}
Figure \ref{fig:decohPlotsB1} shows a comparison of the capacity regions for systems with $B=1$, for three and ten links and different decoherence rates. For all cases, $\mu$ is set to one: for qualitative results, we only need to concern ourselves with the value of $\alpha$ \emph{relative} to $\mu$. In real scenarios, we expect $\alpha$ to be at least one order of magnitude less than $\mu$. From numerical results, we observe that the effect of decoherence on the capacity region is not significant, especially as the number of links grows.
Analysis supports this observation, since we can show that
\begin{align*}
\lim\limits_{k\to\infty}\frac{C_2^{\prime}}{C_2} =1 \text{~and~} \lim\limits_{k\to\infty}\frac{C_3^{\prime}}{C_3} =1.
\end{align*}
\begin{figure}
\centering
\subfloat[][$k=3$]{
 \includegraphics[width=0.49\textwidth]{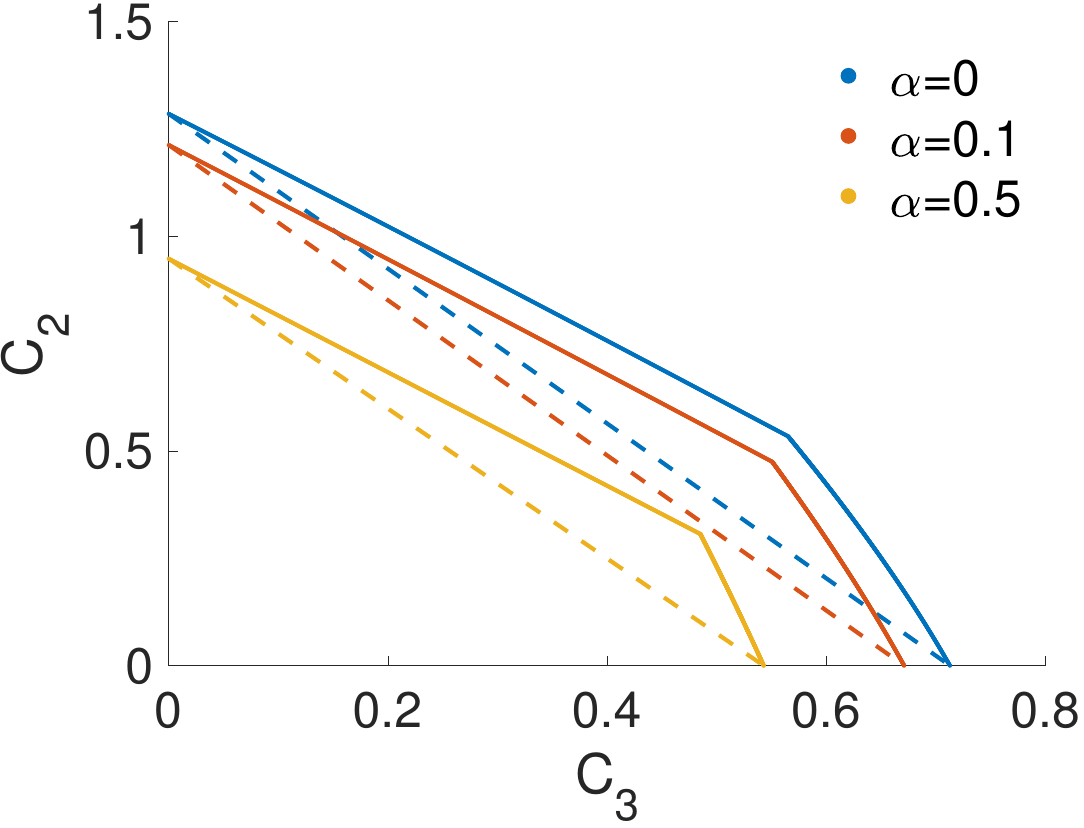}
 }
 \subfloat[][$k=10$]{
\includegraphics[width=0.49\textwidth]{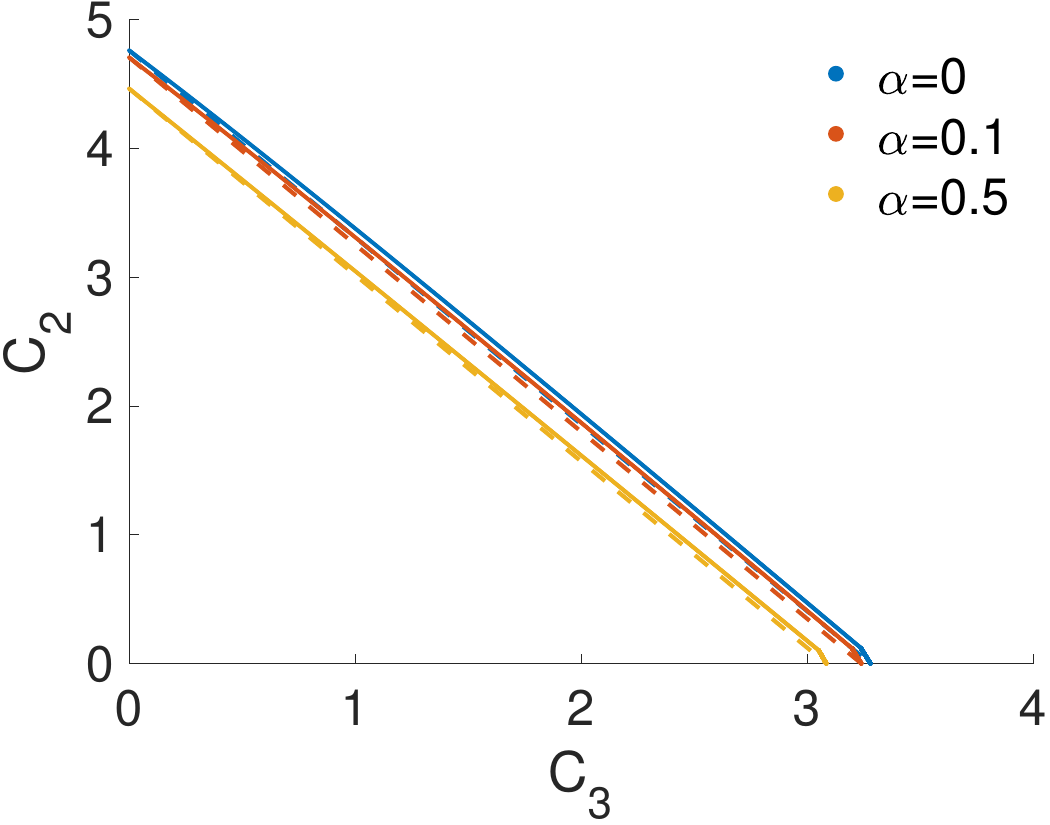}
 }
 \caption{Capacity region for a system of buffer size two and varying number of links $k$, decoherence/storage cut-off rates $\alpha$, and entanglement generation rate $\mu=1$. The solid lines are the upper boundaries of the capacity region, and the dashed are TDM lines.}
 \label{fig:decohPlotsB2}
\end{figure}
Figure \ref{fig:decohPlotsB2} shows a comparison for systems with $B=2$ and varying number of links and decoherence rates. Results are consistent with that of the case $B=1$: the effects of decoherence on capacity are less apparent for larger $k$ values.

%% file: Conclusion_extend.tex
In this work, we explored a set of policies for a quantum switch that can store up to two qubits per link and whose objective is to perform bipartite and tripartite joint measurements to distribute two and three qubit entanglement to pairs and triples of users. We presented analytical results for the case where the per-link buffer has size one. 
We presented a class of policies that achieve a larger capacity region than time-division multiplexing, but found
that as the number of links grows, the advantage of using such policies diminishes. 
We also compared the capacity regions for systems with different per-link buffer sizes and observed that systems with fewer links benefit more from the extra storage space than systems with a larger number of links. Finally, we modeled decoherence and associated storage cut-off times for both types of systems and presented analytical results for the case with per-link buffer size one. Observations and analysis showed that as the number of links increases, the effects of decoherence and storage cut-offs become less apparent on systems.